\documentclass[prd,showpacs,aps,amssymb]{revtex4}
\usepackage{graphicx,epsfig} 
\usepackage{bm}

\newcommand{\beq}{\begin{equation}}
\newcommand{\eeq}{\end{equation}}
\newcommand{\ba}{\begin{array}}
\newcommand{\ea}{\end{array}}
\newcommand{\lsim}   {\mathrel{\mathop{\kern 0pt \rlap
  {\raise.2ex\hbox{$<$}}}
  \lower.9ex\hbox{\kern-.190em $\sim$}}}
\newcommand{\gsim}   {\mathrel{\mathop{\kern 0pt \rlap
  {\raise.2ex\hbox{$>$}}}
\lower.9ex\hbox{\kern-.190em $\sim$}}}

\begin{document}

\title{Signatures of the transition from galactic to extragalactic
  cosmic rays} 

\author{Roberto Aloisio and Veniamin Berezinsky}
 \affiliation{INFN, Laboratori Nazionali del Gran Sasso, I-67010
  Assergi (AQ), Italy}

\author{Pasquale Blasi}
 \affiliation{INAF, Osservatorio Astrofisico di Arcetri, Largo E. Fermi,
  5 - 50125 Firenze, Italy\\
INFN, Laboratori Nazionali del Gran Sasso, I-67010 Assergi (AQ), Italy}
\author{Sergey Ostapchenko}
\affiliation{Institut f\"ur Experimentelle Kernphysik,
University of Karlsruhe, 76131 Karlsruhe, Germany\\
D.V.~Skobeltsyn Institute of Nuclear Physics, Moscow State University,
 119992 Moscow, Russia} 

\date{\today}

\begin{abstract}
We discuss the signatures of the transition from galactic to
extragalactic cosmic rays in different scenarios, giving most
attention to the dip scenario.
The dip is a feature in the diffuse spectrum of ultra-high
energy (UHE) protons in the energy range $1\times 10^{18} - 4\times
10^{19}$ eV, which is caused by electron-positron pair production on
the cosmic microwave background (CMB) radiation.
The dip scenario provides a simple physical description of the
transition from galactic to extragalactic cosmic rays. Here we
summarize the signatures of the pair production dip model for the
transition, most notably the spectrum, the anisotropy and the chemical
composition. The main focus of our work is however on the description
of the features that arise in the elongation rate and in the
distribution of the depths of shower maximum $X_{\rm max}$ in the dip 
scenario. We find that the curve for $X_{\max}(E)$ shows a 
sharp increase with energy, which reflects a sharp
transition from an iron dominated flux at low energies to a proton 
dominated flux at $E\sim 10^{18}$ eV.  
We also discuss in detail the shape of the $X_{\max}$ distributions
for cosmic rays of given energy and demonstrate that this represents a
powerful tool to discriminate between the dip scenario and other
possible models of the transition. 
\end{abstract}

\vskip0.8cm
\pacs{12.60.Jv, 95.35.+d, 98.35.Gi}
\maketitle

\section{Introduction}
\label{introduction}

The observed spectrum of cosmic rays (CR) has a power-law shape at
energies between $E \sim 10^{10}$~eV and $E \sim 10^{15}$~eV, while
several features are observed at higher energies. The {\em  knee} in
the all-particle spectrum consists of a steepening of the power law
behaviour from $E^{-2.7}$ to $E^{-3.1}$. This feature coincides with
the knee in the proton spectrum, but the latter is more pronounced than
the knee in the all-particle spectrum and might be related to a cutoff
in the proton spectrum associated with the maximum energy of
accelerated protons at the sources. The knees in the spectra of
heavier nuclei are found at larger energies but they are not measured
as yet with the same level of accuracy. These knees do not reveal  
themselves as any particular feature in the all-particle spectrum.  

At energies $E_{\rm 2kn} \approx (4 - 8)\times 10^{17}$~eV  
a weak spectral steepening is observed  by the Akeno,    
Yakutsk, Fly's Eye and HiRes detectors. This faint feature is 
referred to as the {\em second knee}. At energy 
$E_a \approx 1\times 10^{19}$~eV a very pronounced flattening of the
spectrum, called {\em ankle} appears. This feature was first
discovered by the Haverah Park detector in the end of '70s. It is now 
seen by most experiments, although the energy where
the ankle is observed depends on the method of analysis adopted for
the spectral reconstruction and is affected by systematic errors
in the energy determination.

Extrapolating the spectrum from higher to lower energies, one
finds the beginning of the ankle at energy $E_a \sim 1\times
10^{19}$~eV. The HiRes collaboration defined the ankle as the
intersection of two power-law spectra, just below and  just above
$E_a$. The intersection energy found in this way is  $E' \approx
5\times 10^{18}$~eV (for a review see \cite{BB}). 

The region between the proton knee and 
the ankle is naturally to be considered as the region where the galactic 
cosmic ray spectrum ends and the extragalactic component begins. 
However, the description of this transition is very model dependent and high
quality observational data are needed in order to discriminate among 
different models.

\subsection{Standard model of galactic cosmic rays}

The standard model for the origin of cosmic rays in the lower
energy part of this transition region is based on the {\it supernova
paradigm}: young supernova remnants (SNRs) may provide the observed
energy density $\omega_{\rm cr} \sim 1\times 10^{-12}~{\rm erg/cm}^3$  
of the galactic cosmic rays and accelerate particles up to a maximum
energy $E_{\rm max} \sim (1-3)\times 10^{15}$~eV for protons (higher
by a factor $Z$ for nuclei with charge $Z$) \cite{Berezhko}. 
Particle acceleration
takes place through first order Fermi acceleration at the supernova
shock. The highest energies mentioned above are reachable only if the
magnetic field in the shock proximity is amplified by a factor
$100-1000$ with respect to the interstellar field, and is rearranged
topologically in order to lead to particle scattering at approximately
the Bohm limit \cite{caprioli}. Magnetic field amplification roughly
to this level has been observed in X-rays \cite{warren} and can be
explained in terms of streaming instability induced by cosmic rays
\cite{bell}, although alternative models of instability cannot be
excluded at the present time. The process of particle acceleration in
the presence of dynamical reaction of the accelerated particles and
magnetic field amplification has been studied recently in
\cite{amato1,amato2,caprioli}. Phenomenological descriptions of the
acceleration process and interesting consequences have recently been
investigated in \cite{Ptuskin}, among other papers. A model  
of the effects of acceleration in SNRs on the overall spectrum of
cosmic rays observed at the Earth has been presented in \cite{BV}.  

  The amplification of the magnetic field takes place in a complex
  chain of nonlinear effects: particle acceleration becomes efficient
  when the field is amplified but streaming instability occurs fast
  enough when particles are accelerated effectively \cite{bell}. This
  situation evolves into a self-regulating nonlinear system. 

  As discussed in \cite{lc83}, the maximum energy achieved by
  particles grows with time during the free expansion phase, but
  saturates at the beginning of the Sedov phase: particles injected at
  the beginning of the free expansion phase or at the beginning of
  the Sedov phase basically reach the same maximum energy, thereby
  confirming that the most important stage for particle acceleration
  in SNRs is the initial part of the Sedov phase. During the Sedov
  phase the shell slows down, and the maximum energy at a given age
  $t$ of the remnant decreases as a consequence of the lack of
  confinement in the shock region of particles accelerated to larger
  energies at previous times. Moreover the effectiveness of magnetic
  field amplification decreases. This situation leads to an
  interesting situation: particles with energy in a narrow range
  around $E_{max}(t)$ escape from the upstream region, with a spectrum
  that at given time is roughly a delta function around
  $E_{max}(t)$. The position of the delta function decreases in energy
  while time progresses. At the same time lower energy particles keep
  being accelerated and stay within the shock. These particles will
  escape the SNR only at much later times. The flux of cosmic rays
  injected by SNRs is the superposition of the flux of particles
  escaping from upstream, integrated over time, and the flux of
  particles accumulated behind the SNR shock and summed over all
  supernova events. In the classical theory of particle acceleration
  the former contribution is unimportant because the spectrum of
  accelerated particles is always steeper than $E^{-2}$ and the total
  energy carried by particles with $E\sim E_{max}(t)$ is negligible. 
  In modern nonlinear theories of particle acceleration at shocks this
  is not the case: the spectra in the highest energy region are
  flatter than $E^{-2}$ and particles with $E\sim E_{max}(t)$ carry
  away from the shock an appreciable amount of energy (e.g. the shock
  becomes radiative). In Ref. \cite{Ptuskin} the authors show that the
  integration over time of the flux of particles escaping from
  upstream during the Sedov phase sums up to a power law with slope
  $\sim 2$. In \cite{BV} the contribution of the particles confined in
  the shock region is calculated in the context of nonlinear theory.

The spectra of different nuclei calculated in \cite{Berezhko,BV} agree 
well with observations of ATIC, JACEE and KASCADE, with the maximum
energy being rigidity dependent $E_{\rm max} \approx 2Z \times
10^{15}$~eV, where Z is the charge of the nucleus. The rigidity dependent
character of $E_{\rm max}$ is the basic feature of this model. At $E
\gsim E_{\rm max}$ the spectra of all nuclei are predicted to have a
sharp cutoff. 

Clearly, these predictions can be compared with observations only after
dressing the {\it standard model} with suitable prescriptions about
the diffusion of cosmic rays in the interstellar medium. With the
standard prescription of diffusion coefficient
$D(E)\propto E^{0.3-0.6}$ the standard model cannot easily explain the
excess of Helium flux below the knee \cite {ant05} and the low level
of anisotropy observed at the knee \cite {agl03,ant04}. We should
however keep in mind that the 
acceleration of helium and other elements in all existing calculations
is carried out in a very phenomenological way, and that the expectations
concerning diffusion are not confirmed in a straightforward way by
more accurate calculations of cosmic ray propagation in the Galaxy
\cite{danielsim}. 

Based on the observation of the proton knee $E_{\rm kn}^p \approx 
(2 - 3)\times 10^{15}$~eV, the end of the galactic cosmic ray spectrum
in the context of the 'standard model' is predicted to coincide with
the iron knee, $E_{\rm kn}^{\rm Fe} \approx (5 - 8)\times
10^{16}$~eV. This is the  
fundamental conclusion of the 'standard model'. If the transition from
galactic to extragalactic CRs occurs at the ankle, $E_a \sim 1\times
10^{19}$~eV, the 'standard model' must be supplemented by additional 
acceleration mechanisms able to boost the maximum energy of the
accelerated particles well above $E_{\rm kn}^{\rm Fe}$. In \cite{BV} 
reacceleration is discussed as a possible mechanism. Since the highest  
energy particles are involved in this process, the chemical
composition at $1\times 10^{17} - 1\times 10^{19}$~eV must be
dominated by iron nuclei.  

\subsection{Extragalactic cosmic rays}
We move now to examining the extragalactic component of cosmic rays.
The traditional model for the transition from galactic to
extragalactic CR is the {\it ankle model} \cite{ankle}. The
attractiveness of this model is provided by its natural character: the
flat extragalactic spectrum crosses the steep galactic spectrum, and
the ankle appears at an energy just above the intersection of the two
components. Another attractive feature of the model is connected with 
the generation spectrum of the extragalactic component which can be as
flat as $E^{-\gamma_g}$ with $\gamma_g\sim 2$. This slope is close to
that predicted by Fermi acceleration at 
non-relativistic shocks ($\gamma=2-2.5$) and at ultra-relativistic
shocks ($\gamma_g =2.2 - 2.3$). It is however important to keep in
mind that these predicted slopes are rather strongly model dependent
in that the spectra can be either flatter, because of the dynamical
reaction of accelerated particles, or steeper, for instance because of
the compression of the magnetic field at the shock surface
\cite{lemsh}. 

  The observed dip at $1\times 10^{18} \leq E \leq 4\times
  10^{19}$~eV can be explained in the context of the ankle model
  following the idea put forward by Hill and Schramm in 1985
  \cite{HS85} in the framework of a two-component model: 
  a steep galactic component encounters a flat extragalactic component
  and produces the dip structure. This idea was later used in the
  calculations of Ref.~\cite{Teshima}.

The drawback of the ankle model resides in its incompatibility with
the 'standard model' illustrated above. Indeed, if iron nuclei start
to disappear at some energy above the iron knee $E_{\rm kn}^{\rm Fe}
\approx (5 - 8)\times 10^{16}$~eV, which particles should fill the gap
between the iron knee and the ankle?  

The pair-production dip model provides an alternative interpretation
of the transition. As has been originally proposed in \cite{BG88}, the
dip can be produced by extragalactic protons with power-law spectrum
due to $e^+e^-$ pair production on CMB photons.
This feature has been studied recently in \cite{BGG,BGG1,us}. 
\begin{figure}[ht]
\begin{minipage}[h]{8cm}
\centering
\includegraphics[width=7.6cm,clip]{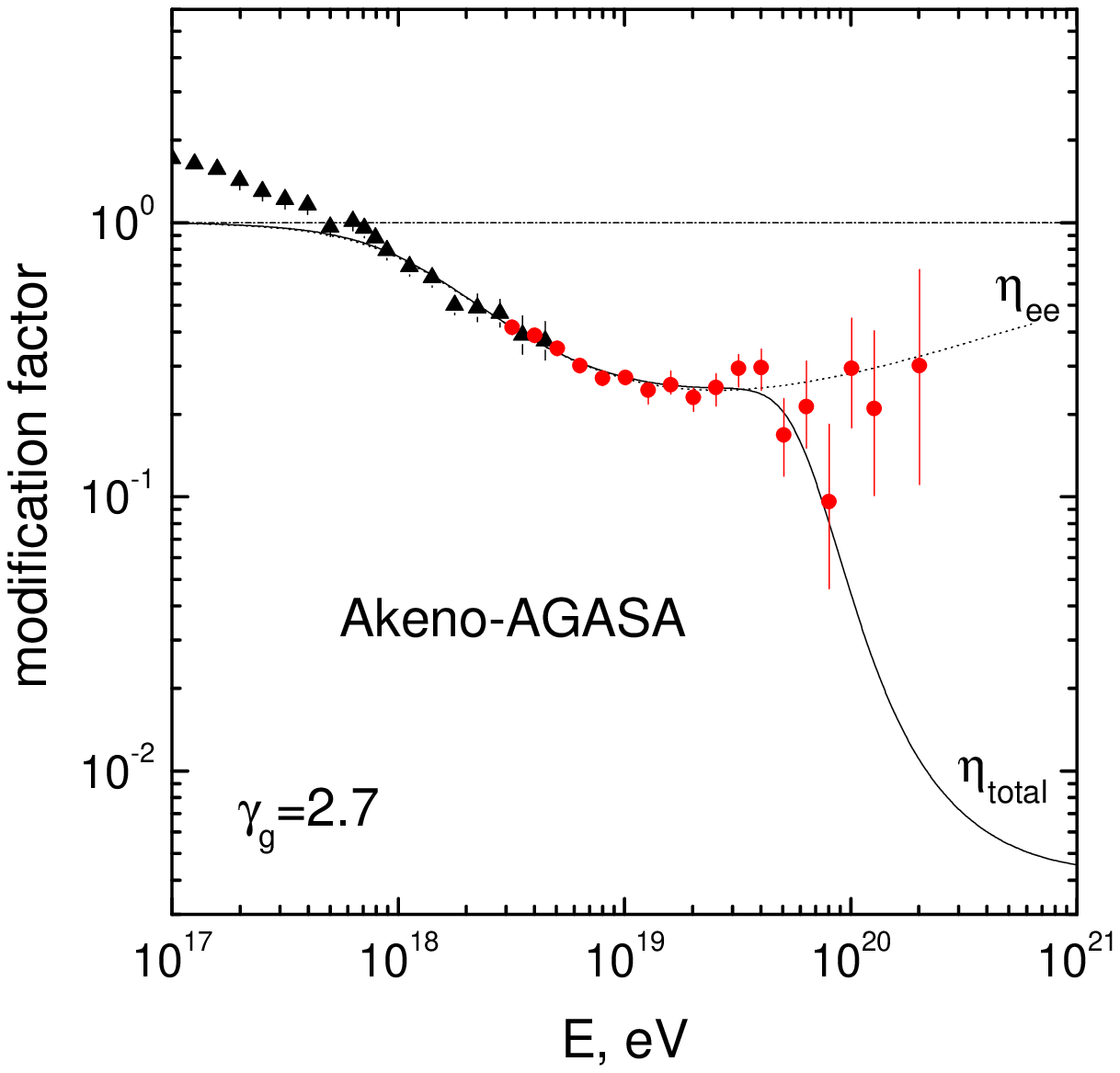}
\end{minipage}
\hspace{2mm}
\begin{minipage}[h]{8cm}
\centering
\includegraphics[width=7.6cm,clip]{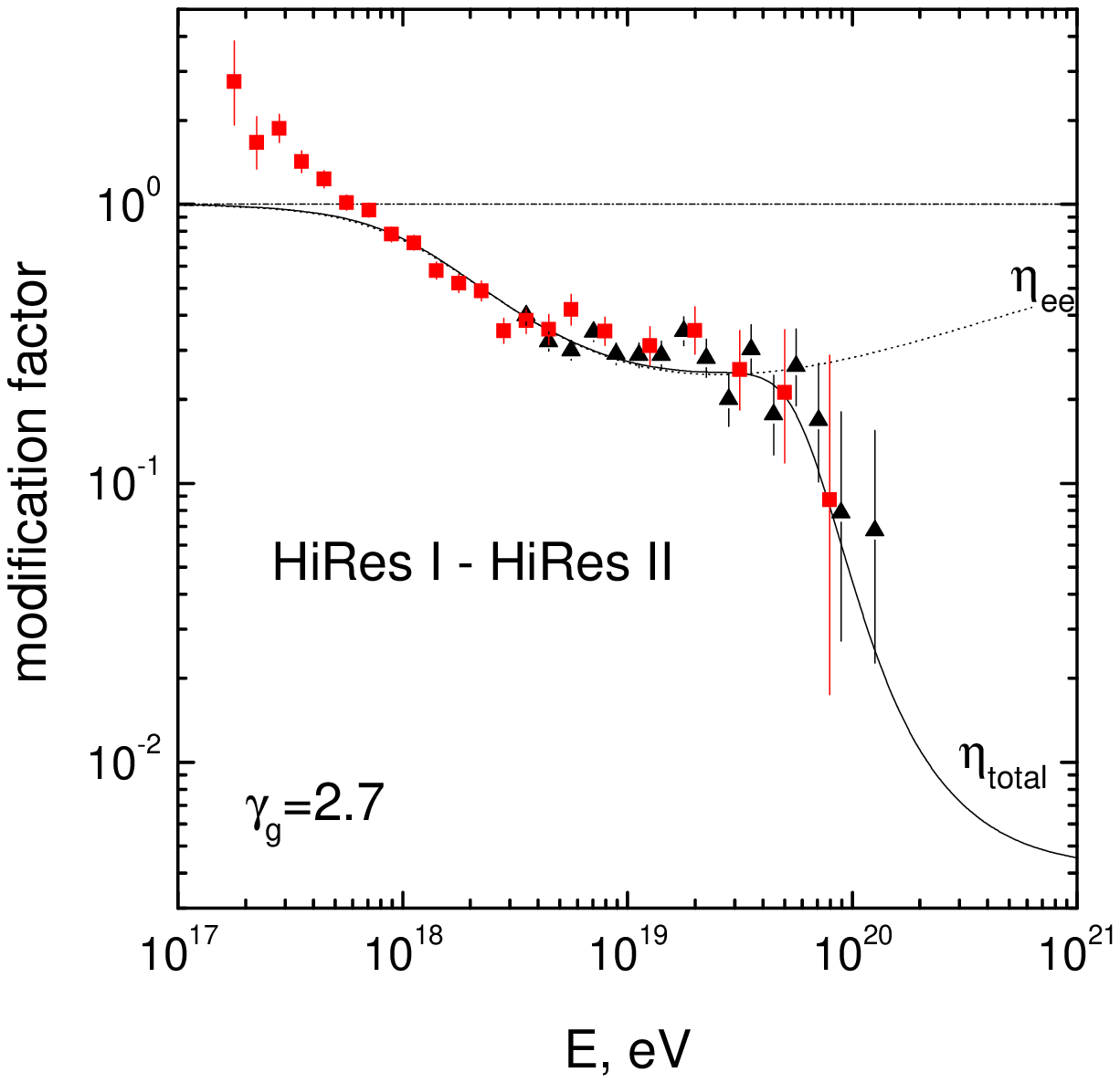}
\end{minipage}
\vspace{2mm}
\begin{minipage}{8cm}
\centering
\includegraphics[width=7.6cm,clip]{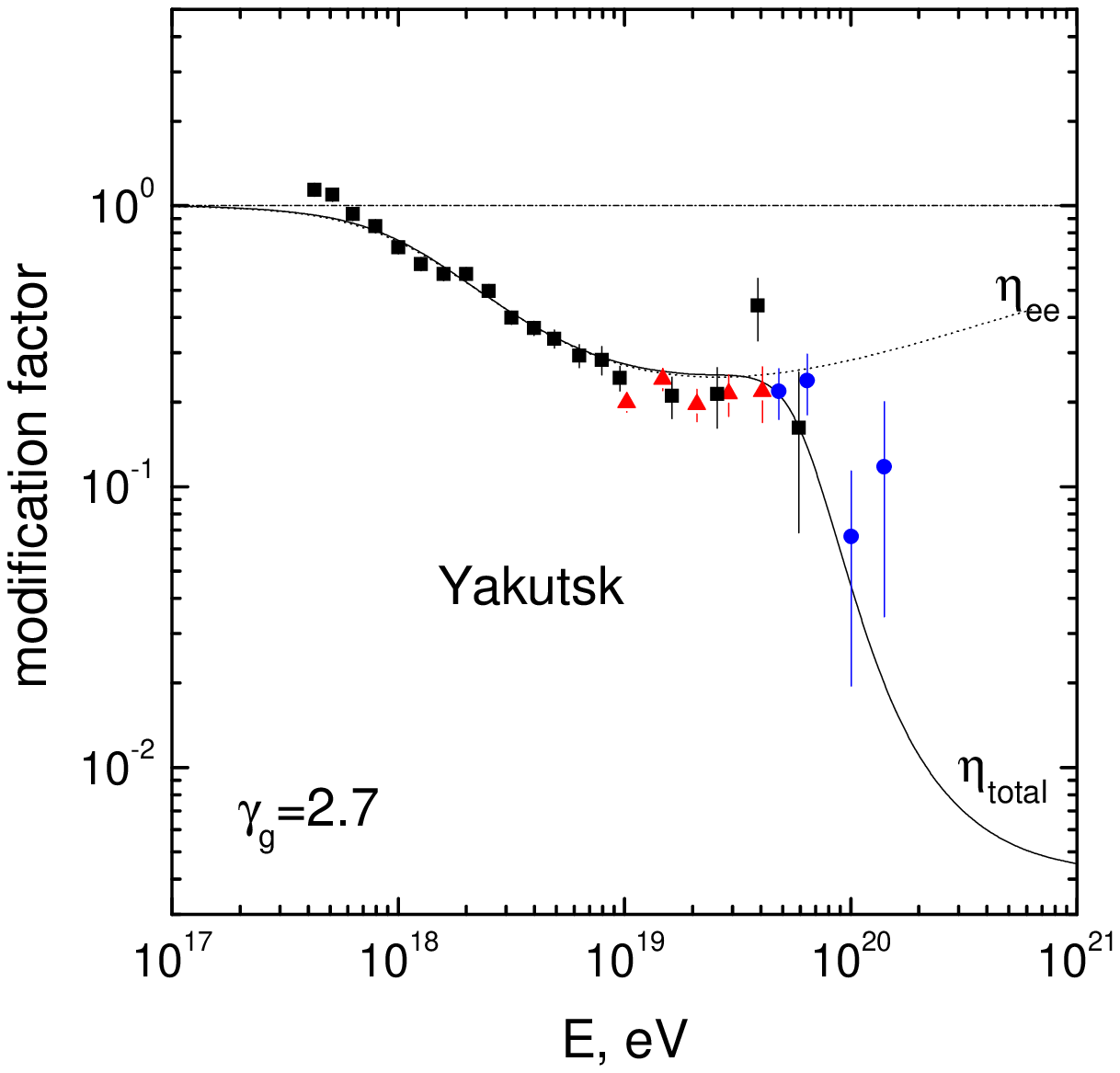}
\end{minipage}
\hspace{7mm}
\begin{minipage}[h]{8cm}
\centering
\includegraphics[width=7.6cm,clip]{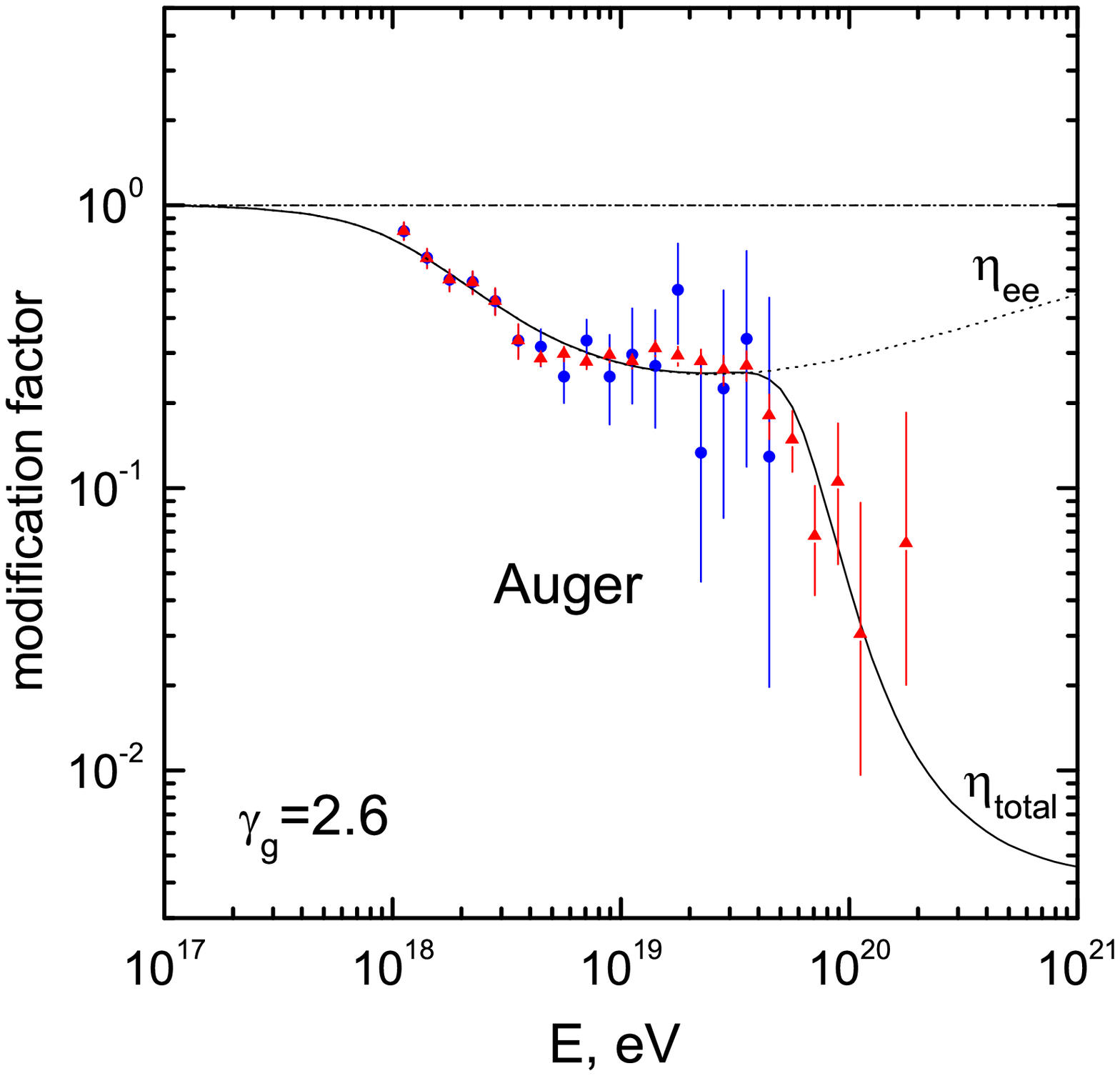}
\end{minipage}
\caption{\label{fig:dips} Predicted dip in comparison with the AGASA 
\cite{agasa}, HiRes \cite{hires}, Yakutsk \cite{Yakutsk} and 
Auger \cite{Perrone} data. The latter are presented as hybrid data, 
shown by circles, and combined data (surface detector data above 4.5
EeV and fluorescence data below), shown by triangles. The comparison of
the dip with Auger data is taken from Ref.~\cite{BGG_ICRC07}.
}
\end{figure}
It is reliably observed in experimental data (see Fig.\ref{fig:dips}),
provided that the generation spectrum is 
$\propto E^{-\gamma_g}$ with $\gamma_g \approx 2.6 - 2.7$. It is
important to stress that this slope refers to the {\it average},
effective spectrum of the sources contained in a shell between
redshifts $z$ and $z+dz$. It can be obtained either by assuming that
all sources contribute the same spectrum $E^{-2.7}$ with a cutoff at
the same maximum energy, or by assuming that single sources contribute
a flatter spectrum (say $E^{-2.3}$) with maximum energies which depend
on the source luminosity and other intrinsic properties
\cite{kach,us}. 

At energies below $E_{\rm cr} \approx 1\times 10^{18}$~eV the
calculated extragalactic spectrum of protons becomes flat, especially
in case of diffusive propagation (see section \ref{sec:dip}), while
the galactic spectrum is very steep ($\propto E^{-3.1}$).
Therefore somewhere below $E_{\rm cr}$ the 
extragalactic spectrum must intersect the steeper ($\propto E^{-3.1}$)
galactic spectrum. The transition occurs at the second knee. The 
prediction of this model -- the strong dominance of proton component at 
$E > E_{\rm cr}$ -- is confirmed by HiRes, HiRes-Mia and Yakutsk data,
while Akeno and Fly's data favor a mixed composition. The dip-based
transition model agrees perfectly with the galactic 'standard model'.  
It is important to notice that the basic ingredient of a transition,
the intersection of a steep galactic spectrum with a flatter
extragalactic one, remains the same in both the dip and the ankle
scenarios. 

An alternative to both the dip scenario and the ankle scenario 
has been put forward in \cite{allard1,allard2}, in which the chemical composition
of the injected extragalactic cosmic rays has been assumed to be 
complex, with a mixture of elements from hydrogen to iron. The 
photo-disintegration of nuclei interacting with IR and CMB radiations 
leads to a spectrum at $E \geq 3\times 10^{18}$~eV  that can fit the observed
all-particle spectrum if an injection spectrum is as flat
as $E^{-\gamma_g}$ with $\gamma_g=2.1-2.3$. A review of the mixed
composition model has been recently presented in \cite{allardlast}.

\subsection{Experimental signatures of the Galactic-Extragalactic transition}
There are basically three types of data which may provide a clue to 
the model for the transition from galactic to extragalactic cosmic
rays. They are {\em spectra, anisotropy} and {\em chemical/mass
composition}.  

The {\em energy spectrum} is the most important source of information
on the transition region, since it is measured with the best accuracy 
in comparison with the other two physical quantities. In general, a
transition from a steep to a flat spectrum is accompanied by a
flattening of the all-particle spectrum. This is certainly true in the
case of the ankle but it does not need to be so in the most general 
case. A typical example is provided by the transition from lighter to
heavier elements around the knee: one might expect a flattening at
each transition, but none is observed in the all-particle KASCADE spectrum.  
In the case of the dip scenario, the transition occurs due to the
intersection of a steep galactic spectrum $\propto E^{-3.1}$ with a 
flat extragalactic spectrum below $1\times 10^{18}$~eV.  But because
of the fact that the transition occurs in a narrow energy range, it
leaves a very weak spectral feature in the all-particle spectrum,
known as second knee. The flatness of the extragalactic spectrum in
the dip model is a general prediction, valid in both cases of straight
line and diffusive propagation. 

The pair-production dip at $1\times 10^{18}\leq E \leq 4\times
10^{19}$~eV is a remarkable spectral feature which characterizes the
transition. It has a very peculiar shape, and its measurement with
high precision may be considered as an evidence of the fact that the
particles detected in this energy region are extragalactic protons
(with at most a small contamination of heavier elements) propagating
through CMB. It is very important that the particle energies measured in 
different experiments operating in this energy region could be
calibrated by the position of the dip. After this calibration the
fluxes measured in different experiments agree with high precision 
and this suggests that the dip is not just an accidental feature in 
the spectrum. This agreement of the dip with the data gives the main 
support of the {\em dip-based model of the transition}.

The third model of transition which is now subject of discussion
is the {\em mixed composition model}. Like the ankle model, it
explains the observed dip in the framework of the Hill-Schramm
two-component model \cite{HS85}. 
The low-energy part of the dip is given by the galactic
component and the high-energy part -- by the extragalactic component
of cosmic rays. The transition occurs at $E \sim 3\times 10^{18}$~eV,
and thus the model agrees well with the 'standard model'. The
injection spectrum required at the sources is compatible with the one
typically expected from diffusive shock acceleration in its basic
version. The mixed-composition model is based on the assumption that the chemical
composition of cosmic rays in extragalactic sources is similar to that
which can be inferred for SNRs after correcting for spallation during
propagation. It is however easy to imagine several astrophysical
situations in which this does not need to be the case. 
Both the ankle model and the mixed composition model are left with the
tough problem of justifying the accidental coincidence of the observed
dip location with the dip generated by pair production, which can be
predicted with high accuracy. 

{\em Anisotropy} may in principle provide information on the
transition: at the transition energy, the anisotropy is expected to
shift from that induced by the location of the Sun in the Galactic
disc to the more isotropic extragalactic cosmic ray flux. A small
anisotropy may be expected in the case of diffusive propagation in the
low energy regime ($10^{17}-10^{18}$ eV), as associated with the
nearest source. The expected anisotropy is however likely to be
undetectable. The anisotropy connected with the galactic sources can 
be detected in the end of the Galactic spectrum (see the discussion in
\cite{BGH}). This possibility is realistic for the ankle transition,
when the maximum energy of the accelerated particles by some additional 
acceleration mechanism may allow particles to reach $1\times
10^{19}$~eV, and the Galactic spectrum cutoff is caused by
insufficient confinement by galactic magnetic field. In this case 
the Galactic protons from a source can reach the observer undergoing
a small deflection angle.   

The {\em chemical composition} gives the most stringent constraint
on the transition models. In the ankle model cosmic rays are expected to
be galactic and iron dominated up to energies in excess of $10^{19}$
eV. In the mixed composition model the transition from galactic to
extragalactic cosmic rays is completed at energies around $3\times
10^{18}$ eV and the chemical composition in this energy region is
mixed. In the dip scenario, the transition is completed at energy
$\sim 1\times 10^{18}$ eV and the composition at this energy is already 
proton-dominated. 

In this paper we concentrate on the signatures of the dip scenario in
terms of the elongation rate and a distribution of shower maximum at given
energy of the primary cosmic rays. We demonstrate that the elongation
rate, irrespectively of the absolute normalization of $X_{\max}(E)$,
which is more model dependent, has in the dip model a {\em sharp
transition} from a composition dominated by iron nuclei to a proton
dominated composition. This sharp transition is absent in the two other
models, ankle and mixed composition, and it may be considered as a
specific signature of the dip model. 

We also calculate the $X_{\max}$ distribution for different energies of
the primaries and propose that the distribution of shower maximum may be
an effective tool to discriminate between the mixed composition model
and the dip scenario. 

The paper is organized as follows: in \S \ref{sec:dip} we summarize
the main predictions of the dip model in terms of the CR spectrum and the 
expected anisotropy. In \ref{sec:elong} we discuss the ankle and the
dip scenarios in terms of the predicted mean elongation rate. The effect
on the distribution of $X_{\max}$ is discussed in \ref{sec:xmax}. We
conclude in \S \ref{sec:discussion}.

\section{The dip model: signatures in the spectrum and anisotropy}
\label{sec:dip}

We start with a short description of the dip-based model of the
transition. 

The pair-produced dip is a faint feature in the spectrum of
extragalactic UHE protons propagating through the CMB. Being
a quite faint feature, the dip is not seen well when the spectrum is
plotted in its basic form, $\log J(E)$ vs $\log E$. The dip appears
more pronounced when it is shown in terms of the {\em modification
factor}, as introduced in \cite{BG88,Stanev00}. The modification
factor is defined as the ratio of the diffuse spectrum $J_p(E)$,
calculated  with all energy losses taken into account, and
the unmodified spectrum $J_p^{\rm unm}$, where only adiabatic energy
losses (red shift) are included: $\eta(E)=J_p(E)/J_p^{\rm unm}(E)$.
The spectrum $J_p(E)$ can be calculated from the conservation of the
number density of particles as
\beq
n_p(E,t_0)dE= \int_{t_{\min}}^{t_0} dt Q_{\rm gen}(E_g,t)dE_g,
\label{conserv}
\eeq
where $n_p(E,t_0)$ is the space density of UHE protons at the
present time, $t_0$, $Q_{\rm gen}(E_g,t)$ is the generation rate per
comoving volume at cosmological time t, and $E_g(E,t)$ is the generation
energy at time t for a proton with energy E at $t=t_0$. This energy
is found from the loss equation $dE/dt=- b(E,t)$, where $b(E,t)$ is
the rate of energy losses at epoch t. The spectrum, Eq. (\ref{conserv}),
calculated for a power-law generation spectrum $\propto E^{-\gamma_g}$
and for a homogeneous distribution of sources, is called
{\em universal spectrum} \cite{BGG}.

Since the injection spectrum $E^{-\gamma_g}$ enters both the numerator
and the denominator of $\eta (E)$, one may expect that the
modification factor depends weakly on $\gamma_g$.

In Fig. \ref{fig:dips} we show the comparison of the modification
factor calculated for $\gamma_g=2.7$ with the observational
data of AGASA, HiRes and Yakutsk, and for Auger data, where
$\gamma_g=2.6$ was used. The presence of the dip in 
the modification factor $\eta_{ee}(E)$ is confirmed by the data at
energies below $E \approx 4\times 10^{19}$~eV. Above this energy the
photopion production dominates (see Fig.~\ref{fig:dips}). Fly's Eye
data, not shown here, confirm the dip equally well. The Auger
spectrum is also in agreement with the dip scenario for
$\gamma_g=2.6$, though with a worse $\chi^2$ .

The dip presented in Fig. \ref{fig:dips} is calculated in terms of the 
universal spectrum, i.e. for a homogeneous distribution of the
sources and assuming no source evolution. In this case we need only
two free parameters for the comparison of the dip with observational
data: $\gamma_g$ and an overall normalization constant (or energy
production rate per unit time and volume -- emissivity
$\mathcal{L}$). For 18 - 22 energy bins in each experiment, the
agreement is characterized by $\chi^2/{\rm d.o.f.} \approx 1$.
In the case of the Auger data $\chi^2/{\rm d.o.f.}$ is larger
\cite{BGG_ICRC07}. 

Despite this impressive agreement with most experimental data, one has
to assess the effect 
of numerous physical effects that may spoil the agreement. 
As was demonstrated in Refs.~\cite{BGG,us}, the inclusion of the 
discreteness in the source distribution, the diffusive propagation of 
protons in magnetic fields (note that the universal spectrum does not
depend on the propagation mode as stated by the propagation theorem
\cite{AB}), and the cosmological evolution with parameters similar to
those observed for active galactic nuclei, do not spoil the
agreement of the dip with the observational data. 
The strong evolution of the sources leads to a flatter injection
spectrum $\gamma_g \approx 2.4 - 2.5$ and to fitting the observed
spectrum at lower energies \cite{BGG} (hep-ph/0204357v1).
The steep generation spectra with $\gamma_g \approx 2.6 - 2.7$,  
source energetics and models of acceleration with low content of nuclei  
are also discussed in Refs.~ \cite{BGG,us}.

The energy calibration of the detectors based upon the position of the
dip provides one more clue to the fact that the agreement with
observations as illustrated in Fig.~\ref{fig:dips} is unlikely to be
accidental. We perform the calibration in the following way: for each
of the three detectors, AGASA, HiRes and Yakutsk, independently, we
allowed for a shift of the energy bins inside the dip by a factor
$\lambda$ to reach the minimum $\chi^2$ in the fit. This procedure
results in $\lambda_{\rm Ag}=0.9$,~  $\lambda_{\rm Hi}=1.2$,~ and
$\lambda_{\rm Ya}=0.9$ for the AGASA, HiRes and Yakutsk detectors,
respectively. After this energy shift the absolute fluxes of all
detectors in the region of the dip and beyond agree with high
precision (see figures in \cite{BGG,us}). 

At  $E\geq 1\times 10^{19}$~eV the dip shows a flattening, which
explains the ankle, seen in the data in Fig.~\ref{fig:dips} at this
energy. We remind again our definition of the ankle as the flat part of
the spectrum (in our case the dip) followed from the high energy
side. One can check from Fig.~\ref{fig:dips} that the beginning of the 
ankle for e.g. HiRes data gives $E_a \approx 1\times 10^{19}$~eV.    

By definition, the modification factor cannot exceed unity.
At energies $E < 1\times 10^{18}$~eV the modification factors of
AGASA-Akeno and HiRes exceed this bound. This signals the appearance
of another component, which is most probably given by galactic cosmic
rays. This is the first indication in favor of a transition from
extragalactic to galactic cosmic rays at $E \sim 1\times 10^{18}$~eV.

\begin{figure}[ht]
  \begin{center}
    \includegraphics[width=14.0cm]{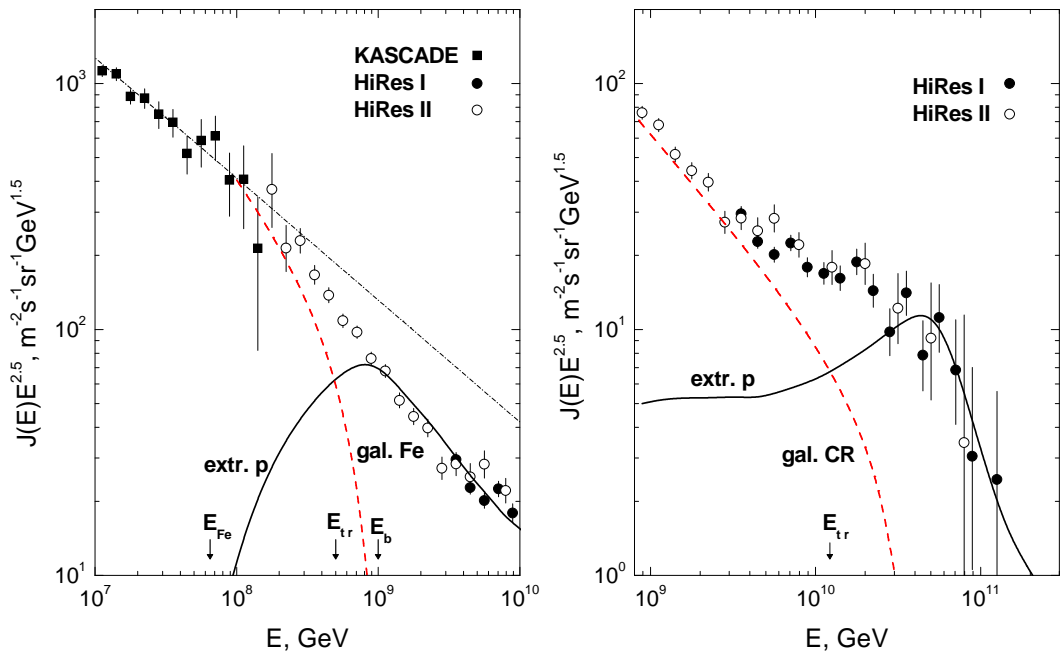}
  \end{center}
  \caption{{\it Left panel: the second-knee transition }. The
    extragalactic proton spectrum is shown for $E^{-2.7}$ generation
    spectrum and for propagation in magnetic field with
    $B_c= 1$~nG and $l_c=1$~Mpc, with the Bohm diffusion at $E \lsim E_c$.
    The distance between sources is $d=50$~Mpc. $E_b=E_{\rm cr}=1\times
    10^{18}$~eV is the beginning of the transition, $E_{\rm Fe}$ is
    the position of the iron knee and $E_{\rm tr}$ is the energy where
    the galactic and
    extragalactic fluxes are equal. The dash-dot line shows the
    power-law extrapolation of the KASCADE spectrum to higher
    energies, which in fact has no physical meaning, because of the
    steepening of the galactic spectrum at $E_{\rm Fe}$.
    {\it Right Panel: the ankle transition},
    for the injection spectrum of extragalactic protons $E^{-2}$.
    In both cases the dashed line is obtained as a result of
    subtracting the extragalactic spectrum from the observed
    all-particle spectrum.}
  \label{fig:transition}
\end{figure}

The transition from galactic to extragalactic cosmic rays in the dip
scenario is displayed in Fig. \ref{fig:transition} (left panel).%
The steep galactic component intersects the flat extragalactic 
proton component, which looks as rising with energy on the graph
because of the multiplication by $E^{2.5}$. This effect is further
strengthened because of the diffusive propagation included in the
calculations.
One can  clearly see the appearance of the second knee (very similar to
the knees observed by KASCADE) that describes this transition. The dashed
line is the inferred galactic cosmic ray spectrum. 

The right panel shows the transition in the traditional ankle model.  

{\em The anisotropy} expected in the dip scenario does not seem to lead to
impressive signatures. At $10^{15}$ eV the observed anisotropy is
small and, if the knee is indeed due to a gradually heavier
composition at higher energies, the anisotropy expected at the iron
knee ($\sim 8\times 10^{16}$ eV) is the same as that of protons at
$3\times 10^{15}$ eV, the proton knee. The second knee defines the
beginning of the transition to extragalactic cosmic rays. At this
energy the composition, in the context of the dip scenario, should
suffer a rather sharp change to a proton dominated one,
which has to  be complete at  $10^{18}$ eV. Extragalactic protons
are most likely isotropic to a large extent: the loss length of
protons in the energy range $10^{17}-10^{18}$ eV is in fact of the
same order of magnitude as the cosmological horizon. In the case of
straight line propagation this distance is certainly larger than the
correlation length which describes the statistical properties of 
gravity-induced clustering of the large scale structure of the
universe. The flux of cosmic rays from a given direction, in this
energy range, is an estimate of the mean density of sources along the
line of sight, which however needs to be very close to the mean
density, since the line of sight extends over an appreciable fraction
of the universe. We conclude that in this case the flux of protons
should be isotropic to a high level. 

In the presence of magnetic field in the intergalactic medium, which
may induce diffusive motion in the low energy region we are interested
in, the issue of anisotropy becomes more complex. As discussed in
several previous works \cite{Lem,AB1}, a magnetic field may induce a
magnetic horizon: if the closest source is at distance $R$ from the
Earth, the propagation time may exceed the age of the universe, in
which case the flux at the energies for which this effect is present
is exponentially suppressed. 

This phenomenon affects the propagation of particles with lower
energies, for which the propagation time is the longest. Assuming that 
particles with energies $10^{17}-10^{18}$ eV manage to reach the Earth
from the closest source, at distance $R$, the flux of cosmic rays is
quasi-isotropic, but not exactly so. In the diffusive regime with
spatial diffusion coefficient $D(E)=\frac{1}{3} \lambda(E) c$, where
$\lambda(E)$ is the energy-dependent pathlength for diffusion, the
anisotropy can be written as 
\begin{equation}
\delta(E) = \frac{I_{\max}-I_{\min}}{I_{\max}+I_{\min}} = \frac{3 D(E)}{c}
\frac{1}{n(E,r)}\frac{\partial n(E,r)}{\partial r},
\end{equation}
where $I(E)$ is the flux of cosmic rays, $n(E,r)$ is the particle
distribution function of cosmic rays at the zero order in the
anisotropy, namely the isotropic component, and $r$ is the distance 
from the source. For a single source, the number density of
particles from the source is $n(r)=\frac{Q(E)}{4\pi r
  D(E)}$. Therefore  
\begin{equation}
\delta = \lambda(E)/R.
\end{equation}
The pathlength $\lambda(E)$ can be related to the power spectrum
$P(k)$ of the fluctuations of the turbulent magnetic field through
\begin{equation}
\lambda(E) = r_L(E) \frac{B_0^2}{\int_{1/r_L(E)}^\infty dk P(k)},
\end{equation}
where $P(k)$ is normalized in a way that $\int_{1/L_0}^\infty dk
P(k)=\eta B_0^2$, with $\eta<1$ being the fraction of the turbulent field
relative to the ordered field $B_0$. For Bohm diffusion $\lambda(E)
= r_L(E)$. For a Kolmogorov spectrum, $P(k)\propto k^{-5/3}$ and one
can show that  
\begin{equation}
\lambda(E) = r_L(E)^{1/3} L_0^{2/3} (1/\eta) = (1/\eta) 0.1 \rm Mpc ~ 
E_{17}^{1/3} B_{-9}^{-1/3} ~ L_{0,Mpc}^{2/3} ,
\label{eq:path}
\end{equation}
where $B_{-9}$ is the strength of the ordered magnetic field in units
of $10^{-9}$ Gauss and $E_{17}$ is the cosmic ray energy in units of
$10^{17}$ eV. At energies somewhat larger than $10^{17}$ eV (for the
reference values of the parameters used here) the propagation rapidly
loses its diffusive character, unless the magnetic field is
unreasonably large (even for $\eta\sim 1$). From
Eq. (\ref{eq:path}) one can also see that in order to obtain that
particles with energy $\sim 10^{18}$ eV suffer the effect of a
propagation time longer than the age of the universe the local
magnetic field must be in the range of a few $10^{-8}$ Gauss. For a 
single source at distance $50$ Mpc, the anisotropy could be of order
$\sim 10^{-3}$ for energies $\sim 10^{18}$ eV. For the case of Bohm
diffusion the anisotropy is easily calculated as $\delta=r_L/R$. For a
source at $50$ Mpc distance one obtains $\delta=2\times 10^{-3} E_{17}
B_{-9}^{-1}$. The numerical value of the expected anisotropy is, not
surprisingly, close to that for Kolmogorov spectrum, since in the
energy region of interest the power spectrum was assumed to reach
saturation (namely the Larmor radius is roughly equal to the size of
the largest eddy).  

These predictions rely however on several assumptions, none of which
appears to be particularly justified. For instance, the density of
sources could be large enough, such that the anisotropy from a single source
is compensated by a spatial distribution of sources. Moreover, even
if the flux reaching the Galaxy is slightly anisotropic, the effect of
the Galactic magnetic field is likely to reduce such anisotropy,
possibly to undetectable levels. 

\section{The elongation rate}
\label{sec:elong}

As discussed in the previous section, in the dip scenario the
transition from galactic to extragalactic cosmic rays occurs sharply
enough, changing from galactic iron to extragalactic protons 
(see left panel of Fig.~\ref{fig:transition}). 
This must result in a steep dependence of the depth of
shower maximum $X_{\max}$ (actually its mean value) as a function of
energy in the range between $10^{17}$ and $10^{18}$ eV. Below
$3\times 10^{17}$ eV we expect $X_{\max}$ being dominated by galactic
iron nuclei. Above $10^{18}$ eV the proton-dominated extragalactic
flux determines the average $X_{\max}$ observed. In this section we
calculate the elongation rates for the dip and ankle models and
compare them with observations.

The results of our benchmark calculations for proton-induced and 
iron-induced showers are shown in Fig. \ref{fig:Xmaxobs} (left
panel):%
\begin{figure}[ht]
  \begin{center}
    \includegraphics[width=8.50cm]{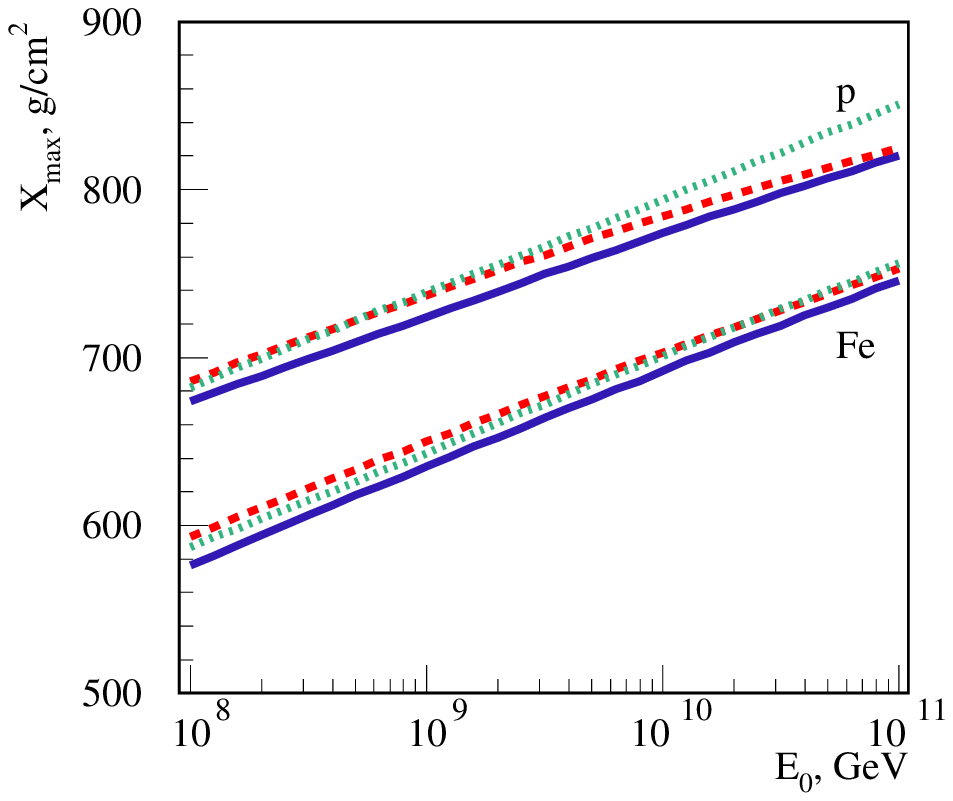}\hspace{5mm}
    \includegraphics[width=8.50cm]{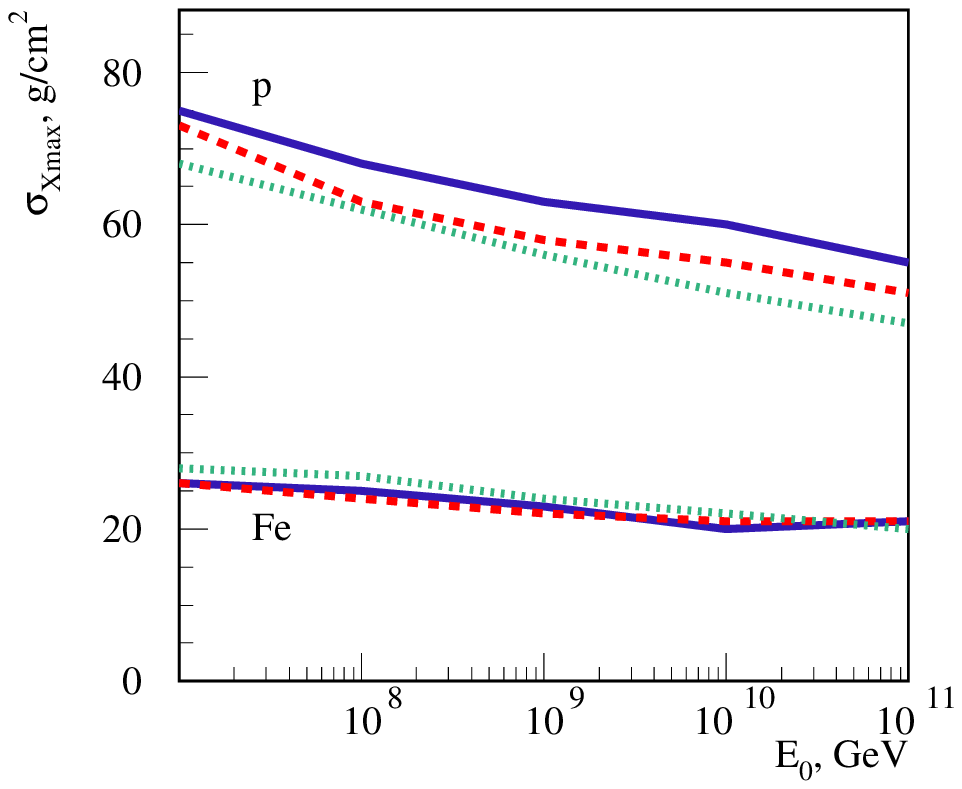}
\end{center}
  \caption{Average penetration depth $\bar X_{\max}$ (left panel) 
    and the variance of  $X_{\max}$ distribution $\sigma_{X_{\max}}$
    (right panel) as functions of energy for protons (upper curves)
    and iron nuclei (lower curves) as calculated using QGSJET,
    QGSJET-II, and SIBYLL models -- solid, dashed, and  dotted lines
    correspondingly. 
  \label{fig:Xmaxobs}}
\end{figure}
we used a standard Extensive Air Shower (EAS) simulation code, CONEX
\cite{berg07}, in order to employ different hadronic interaction
models (here and in the following we simulated 5000 and 1000 showers
per energy for $p$- and Fe-induced EAS correspondingly).  
The solid lines in the Figure refer to QGSJET \cite{qgs}, 
the dashed ones -- to QGSJET-II \cite{qgs2} (version 03), and the
dotted lines -- to SIBYLL 2.1 \cite{sib}. The results of the three
model calculations are within $\sim 20\;\rm g~cm^{-2}$ from each 
other and the  predicted $X_{\max}$ values for proton- and
iron-induced EAS are separated at basically all energies by $\sim
100\;\rm g~cm^{-2}$. As discussed in the next section, the predicted
shower maximum is described by a  distribution whose width varies with
energy (see Fig. \ref{fig:Xmaxobs} (right panel)).  
In the low energy part, around
$10^{17}$ eV, the width of the distribution is $\sim 25\;\rm g~cm^{-2}$
for iron nuclei and $\sim 70\;\rm g~cm^{-2}$ for proton-initiated
showers. These numbers provide a qualitative explanation of the
difficulties in discriminating iron showers from proton-induced ones
(and even more so for elements of intermediate masses). 

Weighing $X_{\max,p}(E)$ and $X_{\max,Fe}(E)$ from
Fig. \ref{fig:Xmaxobs} (left panel) with the flux of cosmic rays in
the form of different chemical components leads to the expected
elongation rate: 
\begin{equation}
X_{\max}(E)=\frac{J_p (E) \bar X_{\max,p}(E)+J_{Fe} (E) \bar
  X_{\max,Fe}(E)}{J_p (E)+J_{Fe} (E)}.
\label{eq:Xmax}
\end{equation}
Here $J_p$ and $J_{Fe}$ are the fluxes of protons and iron
nuclei expected at energy $E$ in a given model. These fluxes take into
account both the galactic contribution and the extragalactic one. In
Eq. (\ref{eq:Xmax}) the quantities $\bar X_{\max,p}(E)$ and $\bar
X_{\max,Fe}(E)$ are those shown in Fig. \ref{fig:Xmaxobs} (left panel).

\begin{figure}[ht]
  \begin{center}
    \includegraphics[width=8.50cm]{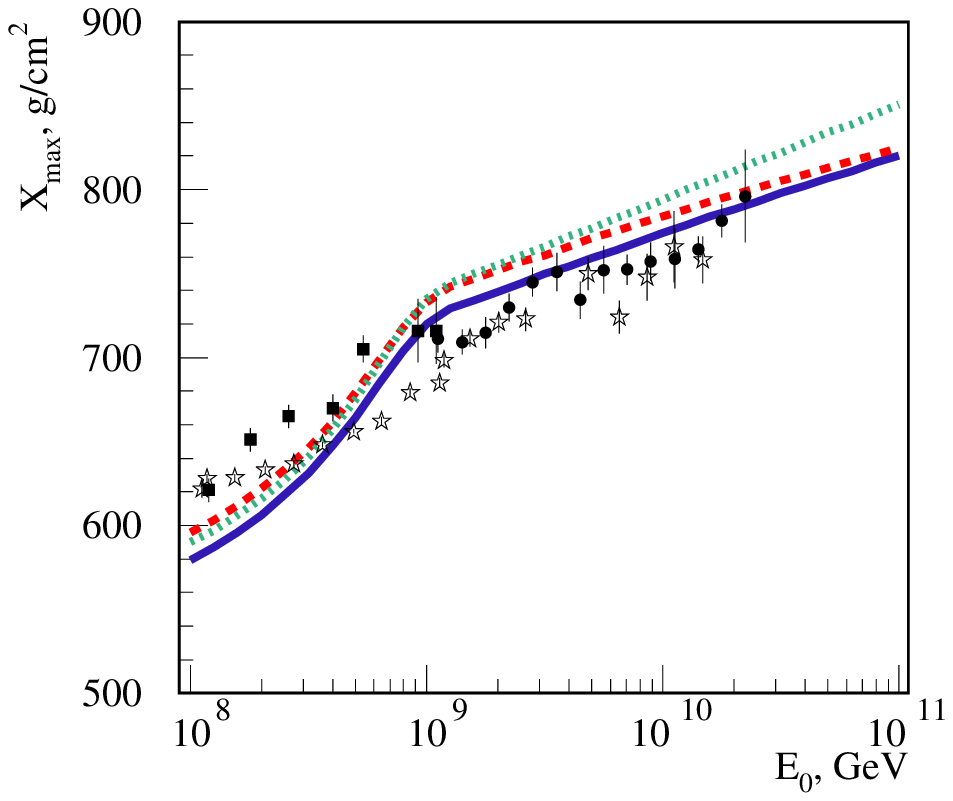}\hspace{5mm}
    \includegraphics[width=8.50cm]{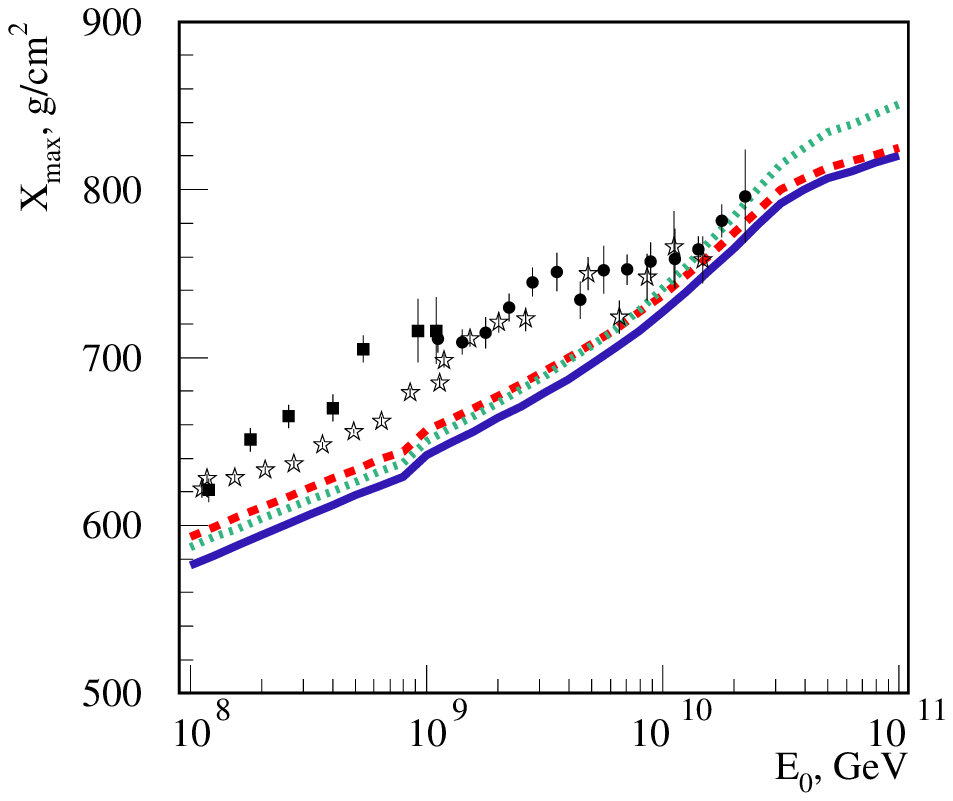}
\end{center}
  \caption{{\it Left panel:} Elongation rate for the dip scenario.
   {\it Right panel:} Elongation rate for the ankle scenario.  The
  three lines, {which presents the calculations are labelled as in
   Fig. 3:  solid, dashed and dotted lines corresponds to QGSJET,
    QGSJET-II, and SIBYLL models, respectively.} 
    The data points are the measurements
  of Fly's Eye (stars) \cite{fly}, HiRes-Mia (squares) \cite{mia} and
 HiRes (circles) \cite{hires-xmax} experiments.}
  \label{fig:dip_ankle}
\end{figure}

In Fig. \ref{fig:dip_ankle}%
we plot the results of our calculations for the penetration depth as a
function of energy for the dip scenario (left panel) and for the ankle
scenario (right panel) in comparison to experimental data of Fly's Eye
\cite{fly}, Hires-Mia \cite{mia} and HiRes \cite{hires-xmax}. 

In the dip scenario (left panel) we identify as a distinctive feature
the sharp rise of the penetration depth at energies between $10^{17}$
eV and $10^{18}$ eV, reflecting the sharp transition from galactic
iron to extragalactic proton-dominated flux. In the calculations
presented here we used Bohm diffusion at energies below 
$1\times 10^{18}$~eV. The shape of $X_{\rm max}(E)$ in the range of
energies considered here remains the same for Kolmogorov 
diffusion, but it becomes smoother for rectilinear propagation of
protons or for very small distances between the sources. The
transition is 
completed at $\sim 1\times 10^{18}$ eV with a composition being
strongly dominated by
protons. In this calculation we neglect the possibility of a
small admixture of nuclei in the extragalactic flux as allowed by the
dip model. In case of 10 - 20 \% admixture of He, the
presented elongation curves change only slightly. Taking into account
a typical systematic uncertainty in the determination of $X_{\rm max}$   
as $20$~- $25$~g/cm$^2$ \cite{mia}, the data plotted in the left panel agree
reasonably well with the dip prediction, especially in the case of the
QGSJET model, and the steep rise of the  elongation rate at $1\times
10^{17} - 1\times 10^{18}$~eV does not contradict the experimental
data.  
In the case of the ankle model, the transition is much smoother in
terms of the chemical composition (right panel), the latter 
becoming proton-dominated only at energies above $10^{19}$ eV. 
In the energy range $(1 - 5)\times 10^{19}$~eV the disagreement  with 
the data exceeds the systematic error in $X_{\rm max}$.

\begin{figure}[ht]
  \begin{center}
    \includegraphics[width=8.50cm]{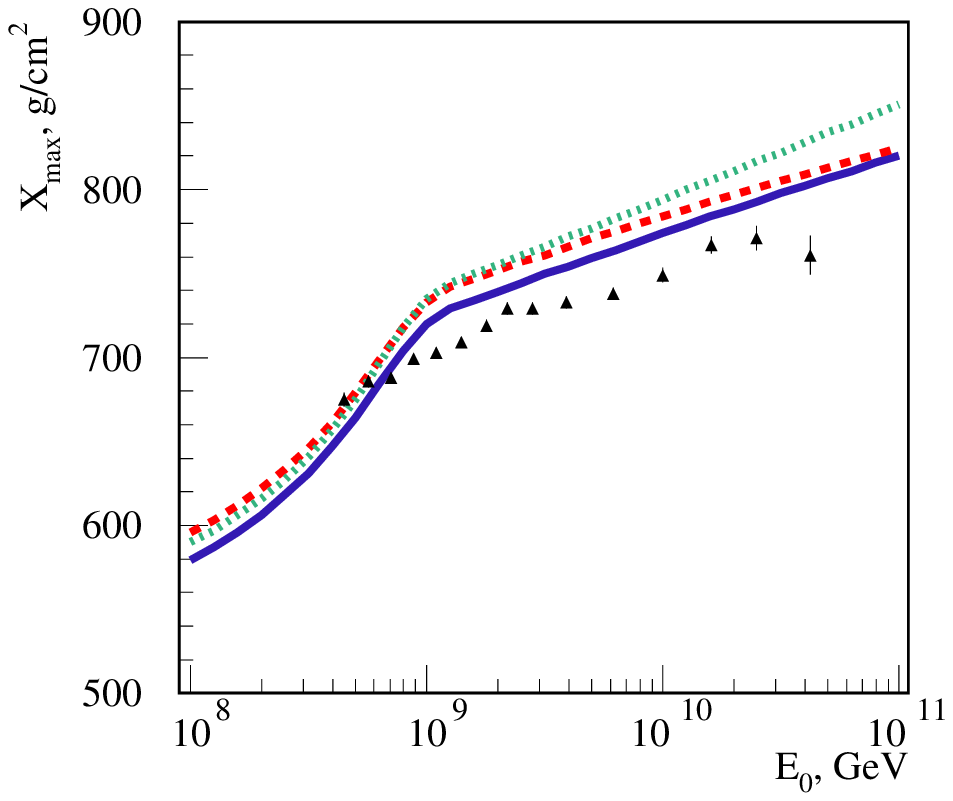}\hspace{5mm}
    \includegraphics[width=8.50cm]{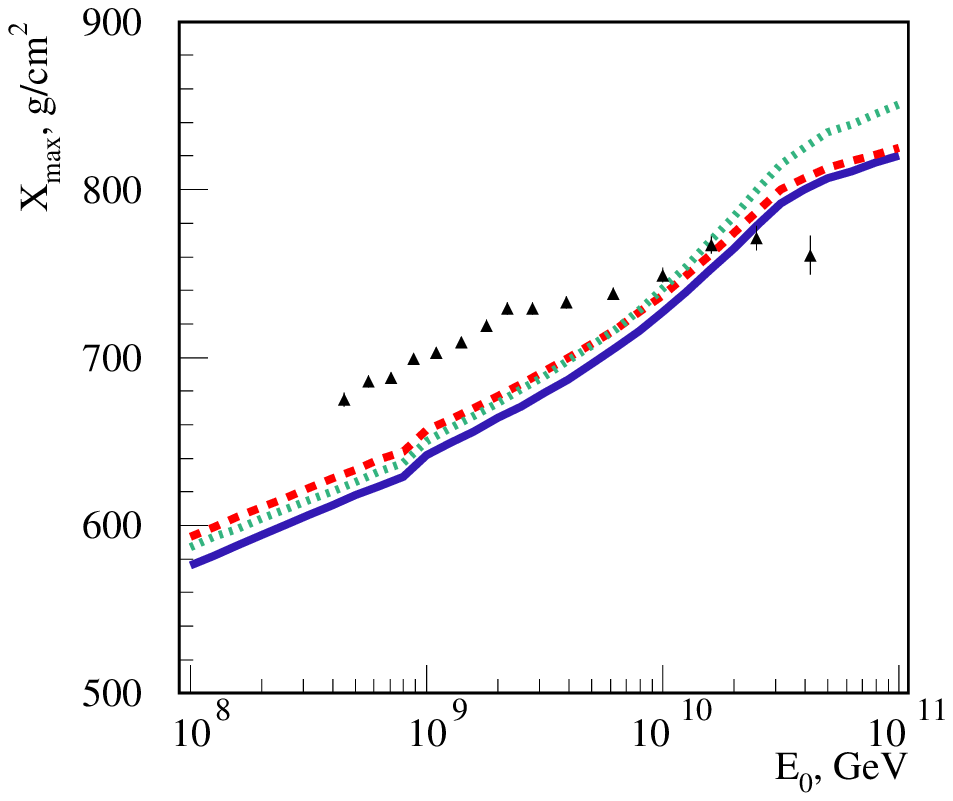}
\end{center}
  \caption{ Elongation rates for the dip scenario
   (right panel) and for that of  ankle (left panel) in comparison 
   with the Auger data \cite{auger-xmax}. The
  three lines are labelled as in Fig. 3. }
  \label{fig:dip_ankle-auger}
\end{figure}

The comparison with the recent Auger data \cite{auger-xmax} is
illustrated separately in Fig.~\ref{fig:dip_ankle-auger}. 
For the dip model (left panel) the disagreement does not exceed
23 $g~cm^{-2}$, if we exclude the highest energy data point.

For the ankle model this disagreement reaches $\sim 60$~g/cm$^2$ in the 
energy range $(5 - 20)\times 10^{17}$~eV.
In principle, in  models which assume a rigidity-dependent
Galactic CR acceleration or propagation one may expect some admixture of
silicon or even lighter nuclei around $10^{17}$ eV (see,
e.g. \cite{hil06}), which rapidly disappear at higher 
energies. Depending on the relative abundance of such lighter elements,
the predicted $X_{\rm max}$ in the left panels of Figs. \ref{fig:dip_ankle} 
and \ref{fig:dip_ankle-auger} may be slightly shifted upwards in the
lowest energy bins, while the corresponding energy dependence
in the interval $10^{17}-2\cdot 10^{17}$ may flatten -- as the importance
of extragalatic protons is then partly compensated by the
disappearance of galactic nuclei which are lighter than iron. An
analysis of such effects goes beyond the scope of the present paper.

The  case of a mixed composition has been discussed in
\cite{allard1,allard2} and it is intermediate between the two cases of
the dip and the ankle models. The agreement of the calculated elongation
rate with the data is the best among these three models, and the choice of 
a chemical composition at the source always allows one to obtain a
good fit to the observations. As far as Auger data are concerned, 
the mixed composition model agrees with the break in elongation rate
at $2\times 10^{18}$~eV and contradicts the highest energy point in 
Auger measurements.  The authors claim as the main
feature of the model the appearance of a plateau in the elongation
rate, to be searched for in future more precise data.  

\section{The $X_{\max}$ distribution}
\label{sec:xmax}

We want to emphasize here that a more effective tool to assess the 
chemical composition in the transition region is provided by an analysis
of the distribution of the shower maximum, which is more sensitive to the
primary composition than the elongation rates plotted in
Fig. \ref{fig:dip_ankle}.  
\begin{figure}[ht]
  \begin{center}
    \includegraphics[width=14.50cm]{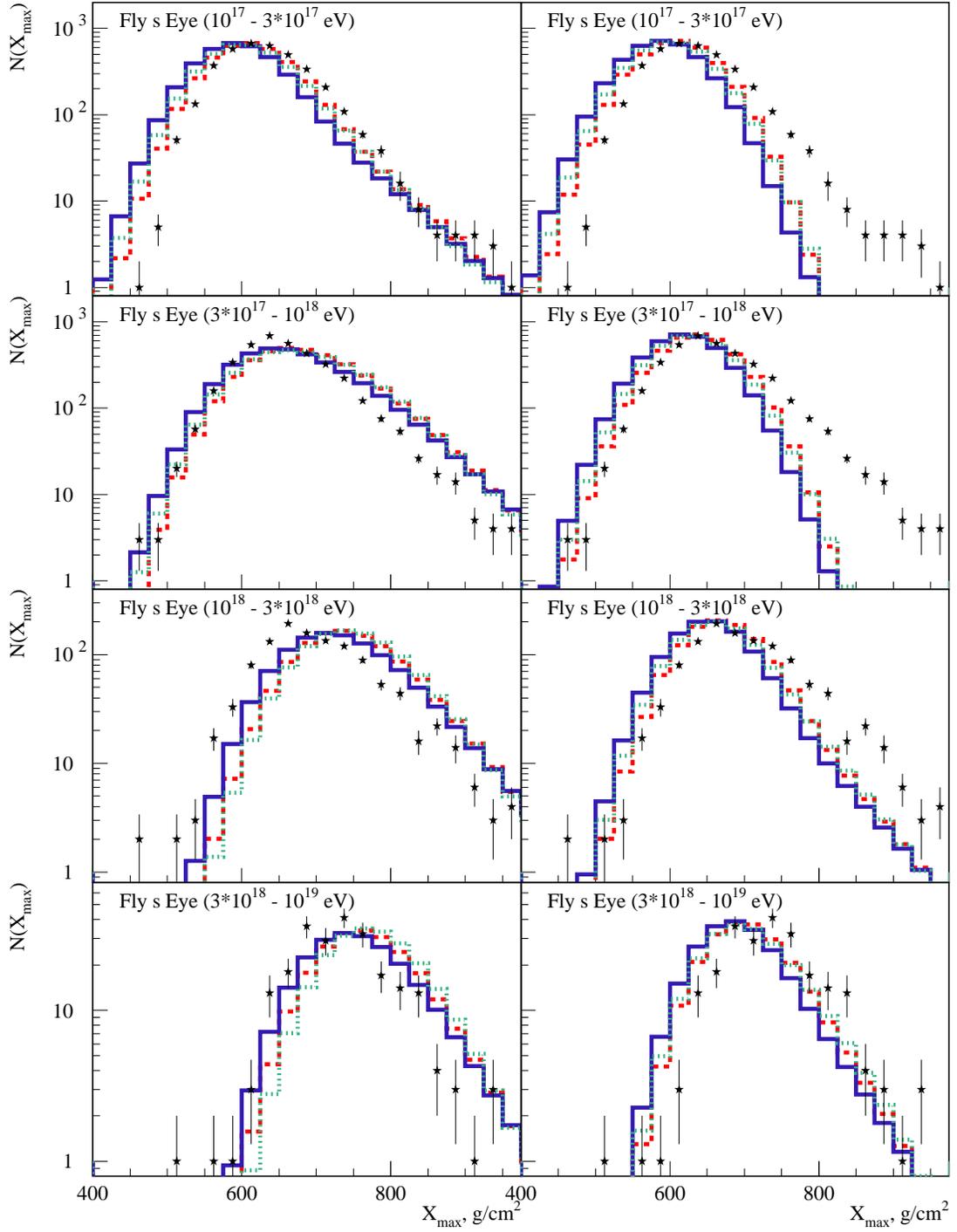}
\end{center}
  \caption{Predicted $X_{\max}$ distribution for the dip scenario (left
  panels) and for the ankle scenario (right panels) in different
  energy bins in comparison with  Fly's Eye data \cite{fly-distr} (points).}
  \label{fig:fly}
\end{figure}
Our benchmark calculation for the distribution of $X_{\max}$ yields the
widths shown in Fig. \ref{fig:Xmaxobs} (right panel), as a function of
the total energy of the nucleus. The results refer to protons (upper
curves) and to iron nuclei (lower curves) for the same interaction
models as discussed in the previous section. It is easy to see that
the model dependence of the calculated $\sigma_{X_{\max}}$ is much
weaker than for the average position of the shower maximum. 
\begin{figure}[ht]
  \begin{center}
    \includegraphics[width=14.50cm]{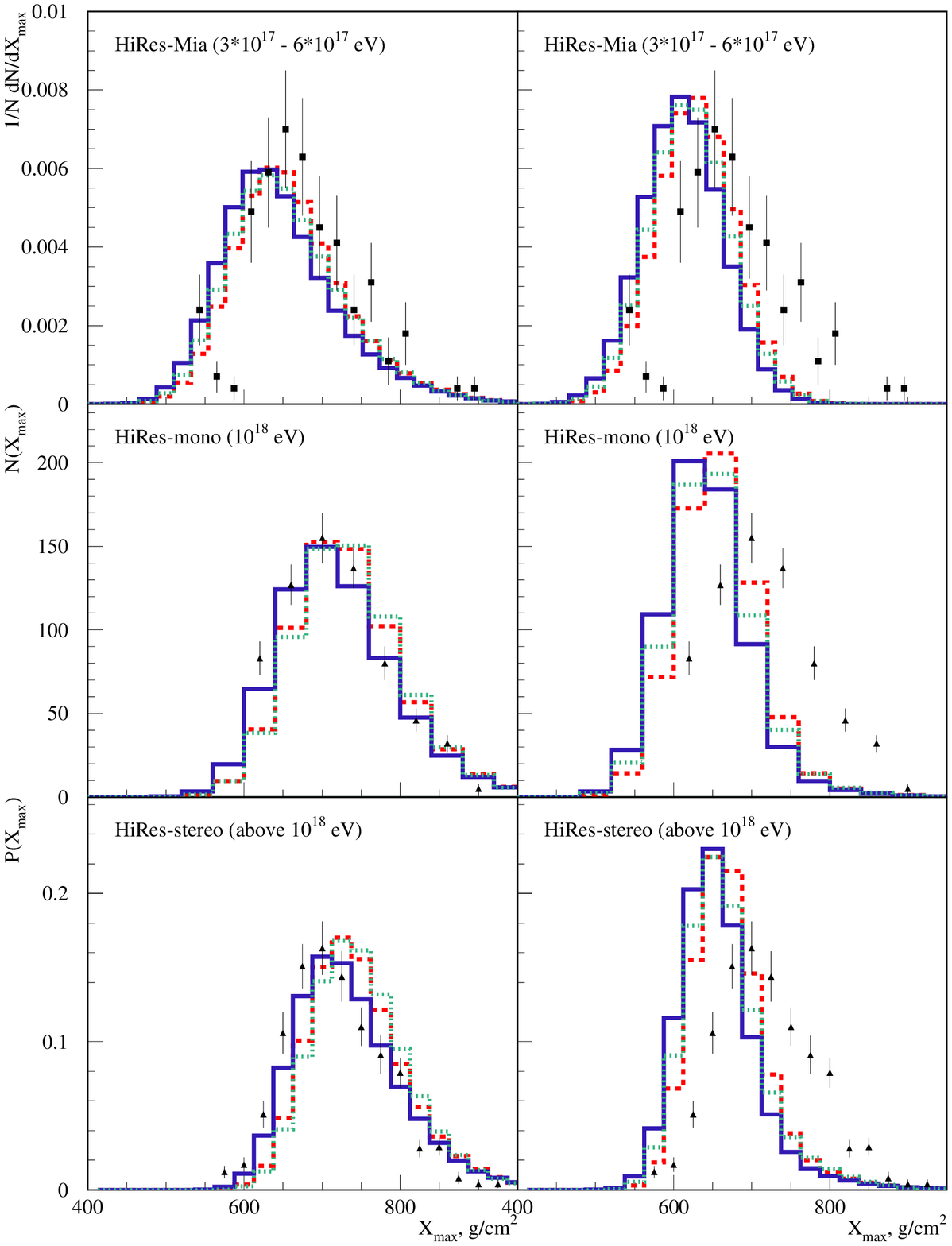}
\end{center}
  \caption{Predicted $X_{\max}$ distribution for the dip scenario (left
  panels) and for the ankle scenario (right panels) in different
  energy bins in comparison with  HiRes data (points).}
  \label{fig:hires}
\end{figure}
For proton-induced EAS the difference in the distribution width is
mainly due  to different total inelastic $\sigma^{\rm inel}_{p-{\rm
air}}$ and diffractive $\sigma^{\rm diffr}_{p-{\rm air}}$ proton-air
cross sections predicted by models \cite{bel06}. It is noteworthy that
the present model differences for $\sigma^{\rm inel}_{p-{\rm air}}$ of
10-15\% will be significantly reduced in the near future, due to the
expected precise measurements of the total proton-proton cross section
at the Large Hadron Collider. In case of primary 
nuclei, the width of the $X_{\max}$ distribution is mainly defined by  
fluctuations of the number of interacting projectile nucleons in
individual nucleus-air collisions \cite{eng92,kal93}, which are
governed by the geometry of the interaction (primarily, by the
variations of the impact parameter of the collision) and are
practically model-independent. Additional model dependence may 
come from the treatment of the fragmentation of the nuclear spectator
part. However, while the two extreme scenarios -- conservation of the
spectator part as a single nuclear fragment or its total break up into
independent nucleons -- give rise to rather different predictions for
EAS fluctuations \cite{kal93}, realistic fragmentation models, being
tuned to the relevant accelerator data, produce very similar results
for $\sigma_{X_{\max}}^{A-{\rm air}}$, as is illustrated by
Fig. \ref{fig:Xmaxobs} (right panel).

The power of using the distribution of penetration depths at given
energy of the primary particle is illustrated in Fig. \ref{fig:fly},%
where we show our results (lines labelled as in the previous section)
compared to the data of the Fly's Eye collaboration \cite{fly-distr}.
The different panels
refer to different energy bins. The left (right) panel presents the
results for the dip (ankle) scenario. To account for the reported
experimental resolution of the shower maximum of $45\;\rm g~cm^{-2}$,
we introduced the corresponding  smearing of the calculated $X_{\max}$
values, using a Gaussian distribution.

In the lowest energy bin ($(1-3)\times 10^{17}$ eV), the shape of the
distribution is well described by the dip model while the fit of the
ankle model seems rather poor. It is in fact interesting to notice
that the tail at depths larger than $\sim 700\rm g~cm^{-2}$ can be
properly fit only if there is an appreciable amount of a light
component. This is the role played by the small fraction of protons in
the left top panel of Fig. \ref{fig:fly}. Moving downwards in
Fig. \ref{fig:fly} corresponds to moving towards larger energies and
the peak of the distributions (for both models) shifts to larger
penetration depths, also due to a lighter mean composition in both
cases.  

In the energy bin $(3-10)\times 10^{17}$ eV, the fit provided by the
dip model still seems acceptable and is definitely better than for the 
ankle model. However, there seems to be a slight excess of the light
component which manifests itself in the tail of the distribution. This could
suggest that a component slightly heavier than protons should be
present. This seems to be confirmed by the plots referring to higher
energies. On the other hand, this effect is more apparent in the
energy bin  $(3-10)\times 10^{17}$ eV, namely where the transition
actually happens in the dip scenario. The exact shape and mix of the
different components in this energy region (galactic plus
extragalactic) is however dependent upon some details, such as the
presence of an extragalactic magnetic field, the possibility of a
solar-wind-like modulation effect due to a galactic wind, which we
have currently no deep insight into. 

In Fig. \ref{fig:hires}%
we show a similar comparison to the data of the HiRes collaboration.
The left (right) column refers to the dip (ankle) model. In the lowest
energy bin ($(3-6)\times 10^{17}$ eV) we compared our results 
with  HiRes-Mia data \cite{mia}. In the middle bin ($E_0\simeq
10^{18}$ eV) we used HiRes mono data \cite{berg04}. In the highest
energy bin ($E_0>10^{18}$ eV) the comparison was made with HiRes 
stereo data \cite{sok05}. Again, a Gaussian smearing of the calculated
$X_{\max}$ values has been introduced according to the reported
experimental resolutions of 45, 41, and $30\;\rm g~cm^{-2}$
respectively. 

The dip scenario fits the data at all energies very nicely, while it
is safe to claim that the ankle scenario does not describe them
correctly. In the energy bin centered at $10^{18}$ eV the peak of the
distribution is already placed at the location expected for proton
showers, as expected for the dip scenario and as already suggested by
the plots on the elongation rate shown in the previous section. In the
highest energy bin, the composition appears to be stabilized to a
proton-dominated one. These conclusions are rather independent of the
interaction model adopted for the calculations. 

\section{Discussion and Conclusions}
\label{sec:discussion}

We discussed the signatures of the transition from galactic to
extragalactic cosmic rays, in terms of spectrum, anisotropy
and chemical composition. Special emphasis has been given to the 
measurement of the elongation rate and to the width of the
distribution of penetration depths $X_{\max}$ in given energy
bins. 

The implications of the different models of the transition for the
{\em spectrum} are profound and in principle the easiest to measure: 
in the ankle scenario the transition occurs at relatively high
energy, $\sim 10^{19}$ eV, as a result of the intersection of a
steep power low galactic component and a flatter extragalactic
spectrum. The ankle scenario is not compatible with the basic
version of the standard model for Galactic cosmic rays, since it
requires a galactic (iron-dominated) component which extends above
$\sim 10^{19}$ eV. 

The dip in the data, as observed by all experiments operating in the
relevant energy region, is naturally explained as being the
pair-production dip. In this case, cosmic rays in the energy region
$10^{18}-10^{19}$ eV are mainly extragalactic protons (with
possibly 10 - 15 \% contamination of nuclei) and the transition between
galactic and extragalactic cosmic rays results in a faint
feature in the all-particle spectrum, known as the second knee. It
represents the lower part of the transition region and occurs, in the
dip scenario, because of the intersection of a steep galactic spectrum
with a flatter extragalactic one. 

In the dip model, the flattening in the spectrum of the extragalactic
component is present both in the case of quasi-rectilinear and
for diffusive propagation. In the latter case the effect may be more
evident, thereby reflecting a flux suppression due to the 
anti-GZK effect and a magnetic horizon \cite{AB1,Lem,us} 
if the magnetic field in the intergalactic medium
is not too small (of order of $0.1 - 1$~nG ). The effect is stronger
in case of Bohm diffusion as compared with Kolmogorov diffusion.

The dip scenario is fully consistent with the SNR paradigm for the
origin of Galactic cosmic rays, according to which Galactic iron
nuclei should be accelerated at most up to $\sim 10^{17}$ eV. 

The pair-production dip fits impressively well the observational data.
When the energy bins of each experiment are shifted to achieve the
minimum $\chi^2$ in comparison with the calculated position of the dip 
(this is what we refer to as the energy calibration of a detector),
the absolute fluxes measured by all experiments agree well with each other. 
This agreement gives another evidence that the spectral coincidence of
the pair-production dip with the data is unlikely to be accidental.

Despite this impressive result, one can fit the data also with a
weighted superposition of different chemical elements at the source,
injected with relatively flat spectra ($\sim E^{-2.3}$). In this mixed
composition scenario, the transition is completed at $\sim 3\times
10^{18}$ eV, thereby being marginally consistent with the basic
predictions of the standard model for the origin of galactic cosmic
rays, based on the SNR paradigm. 

Our predictions on the {\em anisotropy} signal are not exciting: for
the dip model, in both cases of rectilinear(low magnetic
field) and diffusive propagation (larger field) the expected
anisotropy is low and most likely undetectable, especially when the
isotropizing effect of the Galactic magnetic field is taken into
account. These conclusions hold also in the mixed composition model. 
In the ankle scenario, there might be a residual disc anisotropy
associated with the highest energy iron nuclei of galactic origin.  

The most effective tool to infer the nature and location of the
transition is an accurate (and difficult) measurement of the
{\em chemical composition} in the energy region between $10^{17}$ and  
$10^{19}$ eV. Here we discussed the elongation rate and the $X_{\max}$
distribution as two possible tools to gather this information. We also 
compared the predictions for the dip and ankle scenarios with
available data of the  Fly's Eye, HiRes, and Pierre Auger
collaborations. The case of a mixed composition has been
investigated in detail in \cite{allardlast} in terms of the elongation 
rate and was therefore not addressed further here. 

Our benchmark calculations for the penetration depth for proton-
and iron-induced showers have been carried out with SIBYLL, QGSJET
and QGSJET-II hadronic interaction models.
The same interaction codes have been used throughout all
other calculations we carried out. The intrinsic uncertainty in the
mean value of the penetration depth as due to uncertainties in the
interaction models is $\sim 20\;\rm g cm^{-2}$, while the average
separation between proton- and iron-initiated showers as a function of
energy remains of $\sim 100\;\rm g~cm^{-2}$. The distribution of
values of $X_{\max}$ around the mean has a typical width of 
$70\;\rm g~cm^{-2}$ for protons and $25\;\rm g~cm^{-2}$ for iron. This
makes immediately clear why it is particularly hard to nail down the
composition at given energy: only a very large number of showers can
lead to an unambiguous tagging of the composition in terms of the
elongation rate. The task becomes even harder if elements with
intermediate masses between hydrogen and iron are present in
appreciable quantities. 

We calculated the elongation rate expected for the dip and ankle
scenarios. The ankle model provides a bad fit to all sets of
data. The dip scenario is qualitatively much better, but
it still provides only a rough fit to all data sets in agreement
only within systematic energy errors. An exceptional case is given by  
the HiRes data which closely follow the behaviour predicted by the
dip model of the transition. This is also consistent with the original
HiRes claim that the composition becomes proton-dominated already at
$10^{18}$ eV. The general trend observed is that of a transition
from a heavy-dominated composition to a light one in the energy range
between $10^{17}$ eV and a few times $10^{18}$ eV.

The most peculiar prediction of the dip model is that there should be
a sharp transition from heavy to light dominance, starting at the
second knee and ending at $10^{18}$ eV with a proton-dominated
composition. We calculated the elongation rate for this
transition using the most physically justified scenario of diffusive
propagation. In the case of rectilinear propagation the elongation
rate becomes smoother.

The mixed composition scenario leads to a
shallower transition which is completed only at $E\simeq 3\times
10^{18}$ eV. This model seems to provide a better fit to the available
data on the elongation rate (with the possible exception of the HiRes-MIA
results), though the latter show a wide spread which reflects the
inherent experimental systematics. 

We also analyzed the predictions of the dip and ankle models in terms
of the distribution of $X_{\max}$, which is essentially determined by the 
corresponding intrinsic width for a particular type (mass number) of the
primary particle, convoluted with the superposition of the heavy and light
components, as provided by the galactic and extragalactic contributions 
respectively. The calculations have been carried out in energy bins
suitable for the comparison with available data of the Fly's Eye and
HiRes collaborations. 

The lowest energy bin in the Fly's Eye data ($(1-3)\times 10^{17}$ eV)
is very interesting: the comparison of the expected distributions for
the dip and ankle scenarios shows that while the peak of the
distribution in the two cases is essentially at the same position,  
$\sim 600\,\rm g~cm^{-2}$, as expected for iron-dominated showers, the 
tail of the distribution cannot be explained unless a substantial
amount of protons is present, as expected in the dip model. This part
of the distribution cannot be fit by the ankle scenario. 
The dip model also provides a good fit to the Fly's Eye data in the
higher energy bins. The ankle and dip models provide basically the
same distribution of $X_{\max}$ only at energies in excess of
$10^{19}$ eV, where the composition becomes proton-dominated in both
scenarios. 

It is interesting to notice that in the two Fly's Eye data bins that
contain the transition, as expected in the dip scenario ($(3-10)\times
10^{17}$ eV and $(1-3)\times 10^{18}$ eV), the predicted distributions
shows a slight excess of the light component in the tail. This might
suggest that a somewhat heavier component might be needed to improve
the fit. 

The comparison with HiRes data on the distribution of $X_{\max}$ in
the three energy bins $(3-6)\times 10^{17}$ eV (from HiRes-MIA), 
 $E_0\simeq 10^{18}$ eV (from HiRes mono) and $E_0>10^{18}$ eV (from
HiRes stereo) shows a complete agreement with the dip model. The ankle 
model, once more, provides a bad fit to the data.

All these conclusions are very weakly dependent upon the model for
interactions in the atmosphere. 

\section*{ACKNOWLEDGMENTS} 
The work of RA, VB and PB has been partially supported by ASI under
the contract ASI-INAF I/088/06/0 for theoretical studies in High
Energy Astrophysics.  The work of PB was also partially funded through
PRIN 2006.

\end{document}